\documentclass[reprint,superscriptaddress,amsmath,amssymb,aps,10pt,pra,longbibliography,twocolumn]{revtex4-2}
\newcommand{\bra}[1]{\langle #1|}
\newcommand{\ket}[1]{|#1\rangle}

\newcommand{\floor}[1]{\lfloor{#1}\rfloor}
\newcommand{\iddk}{|\mathbbm{1}\rangle\rangle\langle\langle\mathbbm{1}|}

\usepackage{graphicx}
\usepackage{dcolumn}
\usepackage{bm}

\usepackage{tikz}
\usepackage{tikzpagenodes}
\usepackage{tikz-cd}
\usepackage[normalem]{ulem}
\usepackage{dsfont}
\usepackage{amsfonts}
\usepackage{amsmath}
\usepackage{amssymb}
\usepackage{amsthm}
\usepackage{mathtools}
\usepackage{bbm}
\usepackage{bigints}
\usepackage[mathscr]{eucal}
\usepackage[caption=false]{subfig}
\usepackage{ytableau}
\usepackage[none]{hyphenat}
\usepackage{placeins}

\usepackage[hidelinks]{hyperref}
\hypersetup{
    colorlinks,
    linkcolor= {blue!60!black},
    citecolor= {blue!60!black},
    urlcolor = {blue!60!black}
}

\definecolor{yago3}{HTML}{ffb380}

\begin{document}

\title{Exact identification of unknown unitary processes}

\author{Santiago Llorens}
\email{santiago.llorens@ug.edu.pl}
\affiliation{F\'{i}sica Te\`{o}rica: Informaci\'{o} i Fen\`{o}mens Qu\`antics, Universitat Aut\`{o}noma de Barcelona, 08193 Bellaterra (Barcelona), Spain}
\affiliation{Division of Quantum Information, Institute of Informatics, Faculty of Mathematics, Physics and
Informatics, University of Gdańsk, Wita Stwosza 57, 80-308 Gdańsk, Poland}

\author{Arnau Diebra}%
\affiliation{F\'{i}sica Te\`{o}rica: Informaci\'{o} i Fen\`{o}mens Qu\`antics, Universitat Aut\`{o}noma de Barcelona, 08193 Bellaterra (Barcelona), Spain}

\author{Michal Sedl\'ak}
\affiliation{RCQI, Institute of Physics, Slovak Academy of Sciences, Dúbravská cesta 9, 84511 Bratislava, Slovakia}
\affiliation{Faculty of Informatics, Masaryk University, Botanická 68a, 60200 Brno, Czech Republic}

\author{Ramon Muñoz-Tapia}
\affiliation{F\'{i}sica Te\`{o}rica: Informaci\'{o} i Fen\`{o}mens Qu\`antics, Universitat Aut\`{o}noma de Barcelona, 08193 Bellaterra (Barcelona), Spain}


\begin{abstract}
The accurate identification of faulty hardware is a fundamental requirement for reliable quantum information processing. We address this problem in a quantum setting, where a series of $n$ devices is intended to apply the same unitary operation, but $k$ malfunctioning devices among them apply a different, unknown unitary action. Under the assumption of complete ignorance regarding the specific unitary transformation applied, we model our hypotheses using representation-theoretic tools and study the zero-error protocol for identifying these faulty devices. We derive the optimal success probability for the single- and two-anomaly scenarios, demonstrating that it is independent of the total number of devices in the series. Furthermore, we present a simple protocol that makes use of ancillary systems that achieves this optimal limit. Notably, this protocol offers significant operational advantages, such as allowing us to test each device independently. Finally, we extend our analysis to the general scenario in which both the number of anomalies and the local dimension of the systems are arbitrary, evaluating our protocol's performance and conjecturing its global optimality in the general case.

\end{abstract}

\maketitle

\section{Introduction}

The identification of rare events that deviate from expected behavior is known as anomaly detection. It is a fundamental primitive in classical data analysis and signal processing~\cite{tartakovsky_sequential_2014}, with a wide range of applications~\cite{siris_application_2004,thottan_anomaly_2003,ahmed_survey_2016,zipfel_anomaly_2023}. The ability to discern anomalies allows for timely interventions, improving the reliability and efficiency of engineered systems. 

As quantum technologies advance, the need for anomaly identification naturally extends to the field of quantum information. Quantum computing~\cite{nigg_quantum_2014,corcoles_demonstration_2015,arute_quantum_2019}, quantum communication~\cite{khatri_principles_2020,liao_satellite--ground_2017}, and quantum sensing~\cite{degen_quantum_2017,reiter_dissipative_2017} all rely on precise control of unitary evolutions. Deviations from the intended dynamics may arise from imperfect gate implementations, uncontrolled environmental interactions, or even adversarial interventions in quantum networks. Identifying such deviations is therefore essential for the development of robust quantum technologies. 

In this work, we develop optimal protocols for identifying anomalous unitary processes within the framework of quantum testers. We consider a collection of devices that are intended to implement a fixed, known unitary transformation, taken as the reference behavior. Among them, a subset is anomalous in the sense that they implement an unintended unitary, unknown to the observer. Our goal is to determine 
without error which devices are anomalous.
One may also adopt a minimum-error approach, allowing for the possibility that the identification of faulty devices is not always correct. 
In the present context, however, it seems more natural to impose a zero-error condition, ensuring that devices operating as intended are never mistakenly discarded. This formulation also enables us to determine the structure of the optimal protocol and to derive analytical expressions for the success probability. Moreover, the zero-error---or unambiguous approach---has been shown, in related scenarios, to yield the exact leading-order term for the minimum-error protocol in the limit of a large number of devices~\cite{skotiniotis_identification_2024,llorens_quantum_2024}.

We cast the task as a discrimination problem over quantum channels. Employing the Choi–Jamiołkowski isomorphism~\cite{choi_completely_1975,jamiolkowski_linear_1972}, which maps quantum channels with positive operators, we formulate the problem in the quantum testers framework, in which the role of generalized measurements,  given by a Positive Operator Valued Measure (POVM), is played by quantum testers (or strategies)~\cite{chiribella_theoretical_2009}. Since the unitary defining the anomalous devices is unknown, we assume that all unitaries are equally likely. Accordingly, we adopt as figure of merit the average success probability of correctly identifying the positions of the anomalies over all possible unitaries and all possible placements.

There is a substantial increase in complexity when moving from one to two or more anomalies. In the single-anomaly case, the corresponding Choi operators exhibit local symmetries, which allow the optimization to be restricted to a reduced family of testers. Within this restricted set, we derive a closed-form expression for the optimal success probability, valid for arbitrary local dimensions and any number of devices. The resulting protocol admits a simple local and parallel implementation requiring only maximally entangled probe states. A parallel strategy is more direct to implement, as devices are 
all tested simultaneously without adaptive control, in contrast to the more general sequential strategy, where devices are tested in a given order and one can insert intermediate quantum operations between uses that may depend on previous outcomes. 

When two anomalies are present, the local symmetries are lost, and instead a global symmetry involving all parties emerges. Although restricting the optimization to symmetry-preserving subspaces reduces its complexity, the set of feasible testers remains large, making the problem substantially more challenging than in the single-anomaly case. In particular, these testers can be characterized in terms of the algebra of partially transposed permutations~\cite{cox_blocks_2008,grinko_gelfand-tsetlin_2023,balanzo-juando_positive_2024,studzinski_irreducible_2025}.
Nevertheless, we prove that for qubits and parallel implementations, the same local protocol as in the single-anomaly case remains optimal. The proof combines semidefinite programming (SDP) methods with representation-theoretic techniques. In particular, it relies on a block-diagonal decomposition based on mixed Schur--Weyl duality \cite{turaev_operator_1990,koike_decomposition_1989,grinko_mixed_2025}, which reduces the analysis of otherwise high-dimensional operators to tractable components. We further show that the optimal success probability remains nonzero for any number of devices. Moreover, numerical evidence for small system sizes indicates that sequential strategies offer no advantage over parallel ones, suggesting that our results may extend to the fully general setting. 

The analysis for the case of more than two anomalies is qualitatively similar,  but involves some complex
combinatorial quantities.  In the qubit case, $d=2$, these quantities reduce to Catalan numbers, allowing a closed-form expression for an arbitrary number of anomalies. Remarkably, the success probability is finite and independent of the number of devices of the network. Thus, for parallel strategies, we provide the complete solution for arbitrary number of anomalies and devices.

This paper is organized as follows. We begin by introducing the problem and formulating it within the framework of quantum testers. We then cast the task as a SDP, which enables a general analysis of the single-anomaly case and leads to the derivation of the optimal protocol. Next, we study the case of two anomalies, focusing on parallel implementations for qubits, and extend the analysis to an arbitrary number of anomalies and dimension. The paper concludes with a discussion of the results and possible directions for future research.
We also include two technical appendices: the first provides a brief introduction to the algebra of partially transposed permutations, while the second contains a detailed derivation of the success probability for parallel local strategies.

\section{Setting of the problem}

Consider a collection of devices $\mathcal{D}_i$, which are labeled by $i=1,2,\ldots,n$, and each implements some unitary channel $\mathcal{V}(\rho)=V\rho V^\dagger$ acting on a single ($d$-dimensional) quantum system, labeled by $i$. Among them, a subset of $k$ devices is faulty, and apply an unknown anomalous unitary transformation $\mathcal{W}(\rho)=W\rho W^\dagger$, while the remaining devices apply the expected known unitary channel $\mathcal{V}$.

Since the action of the known unitary can be inverted, without loss of generality, we reduce the problem to distinguishing the identity channel $\mathcal{I}(\rho):=\rho$, from the action of the unitary channel
\begin{equation}
    \mathcal{U}(\rho):=U\rho\,U^\dagger\,, \quad\mathrm{with}\hspace{0.2cm} U=V^\dagger W\,.
\end{equation}
For a given subset $r\subset\{1,\ldots,n\}$ of size $|r|=k$, we define the corresponding hypothesis as the channel
\begin{equation}
    \tilde{\mathcal{F}}_r(\cdot) := \mathcal{I}_{\bar{r}}(\cdot) \otimes\mathcal{U}^{\otimes k}_r(\cdot)\,,
\end{equation}
where $\bar{r}$ denotes the complementary set of $r$. That is, every non-faulty device acts trivially, while faulty devices apply the same unknown unitary $U$.

The anomaly identification problem thus consists of distinguishing among the set of channels
\begin{equation}
    \{\tilde{\mathcal{F}}_r:\ r\subset\{1,\ldots,n\}\,,\ |r|=k\}\,,
\end{equation}
corresponding to the $N=\binom{n}{k}$ possible locations of the faulty devices, for the choice of single-shot strategies, in which each device can be tested at most once. We model our complete lack of knowledge about the location of the anomalies by considering all possible locations equally probable, i.e., the priors are $\eta_r=1/N$.

We use the Choi-Jamiołkowski isomorphism and the tester formalism. For a channel $\Phi: \mathcal{H}_\mathrm{in}\to \mathcal{H}_\mathrm{out}$, its Choi matrix is
\begin{equation}
J_\Phi = \sum_{i,j=0}^{d-1} \ket{i}\bra{j} \otimes \Phi(\ket{i}\bra{j}) \in \mathcal{H}_\mathrm{in} \otimes \mathcal{H}_\mathrm{out}\,,
\end{equation}
where ${\ket{i}}$ is the computational basis of a $d$-dimensional Hilbert space. For notational clarity, we use the subindices ``$\mathrm{in}$" and ``$\mathrm{out}$".

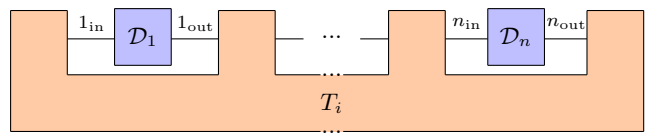
\begin{figure}[ht]
    \centering
    \begin{tikzpicture}[scale=0.5]
            \setlength{\unitlength}{0.1mm}
            \fill[yago3!70] (-8,-0.25) rectangle (8.75,1.25);
            \fill[yago3!70] (-8,1.25) rectangle (-6.5,2.95);
            \fill[yago3!70] (-2.5,1.25) rectangle (-1,2.95);
            \fill[yago3!70] (2,1.25) rectangle (3.5,2.95);
            \fill[yago3!70] (7.25,1.25) rectangle (8.75,2.95);
            \draw[-] (-8,2.95)--(-6.5,2.95);
            \draw[-] (-8,2.95) -- (-8,-0.25);
            \draw[-] (-6.5,2.2) -- (-5.25, 2.2); 
            \draw[-] (-6.5,2.95) -- (-6.5,1.25);
            \draw[-] (-6.5,1.25) -- (-2.5,1.25);
            \draw[-] (-8,-0.25) -- (0.2,-0.25);
            \draw[-] (0.8,-0.25) -- (8.75,-0.25); 
            \draw[-] (-2.5,2.95) -- (-2.5,1.25);
            \draw[-] (-1,2.95) -- (-1,1.25);
            \draw[-] (-2.5,2.95)--(-1,2.95);
            \draw[-] (-1,1.25) -- (0.2,1.25);
            \draw[-] (0.8,1.25) -- (2,1.25);
            \draw[-] (-1,2.2) -- (-0.25,2.2); 
            \draw[-] (-3.75,2.2) -- (-2.5,2.2);
            \draw[-] (1.25,2.2) -- (2,2.2);
            \draw[-] (2,2.95) -- (2,1.25);
            \draw[-] (3.5,2.95) -- (3.5,1.25);
            \draw[-] (2,2.95)--(3.5,2.95);
            \node at (0.5,0.5) {$T_i$};
            \node at (0.5,-0.25) {\text{...}};
            \node at (0.5,1.25) {\text{...}};
            \draw[-] (3.5,1.25) -- (7.25,1.25);
            \draw[-] (3.5,2.2) -- (4.25+0.375,2.2); 
            \draw[-] (5.75,2.2) -- (7.25,2.2); 
            \draw[-] (7.25,2.95) -- (7.25,1.25);
            \draw[-] (8.75,2.95) -- (8.75,-0.25);
            \draw[-] (7.25,2.95)--(8.75,2.95);
            \fill[blue!25] (-5.25,2.5-1.) rectangle (-3.75,4-1.); 
            
            \draw[-] (-5.25,2.5-1.) -- (-3.75,2.5-1.);
            \draw[-] (-5.25,4-1.) -- (-3.75,4-1.);
            \draw[-] (-3.75,2.5-1.) -- (-3.75,4-1.);
            \draw[-] (-5.25,2.5-1.) -- (-5.25,4-1.);

            \node at (-4.5,3.25-1.) {$\mathcal{D}_1$};

            \node at (0.5,3.25-1.) {\text{...}};
            
            \fill[blue!25] (5-0.375,2.5-1.) rectangle (6.5-0.375,4-1.);

            \draw[-] (5-0.375,4-1.) -- (6.5-0.375,4-1.);
            \draw[-] (5-0.375,4-1.) -- (5-0.375,2.5-1.);
            \draw[-] (5-0.375,2.5-1.) -- (6.5-0.375,2.5-1.);
            \draw[-] (6.5-0.375,2.5-1.) -- (6.5-0.375,4-1.);

            \node at (5.375,3.25-1.) {$\mathcal{D}_n$};    
            \node at (-5.85,2.6) {\scriptsize $1_\mathrm{in}$};  
            \node at (-3.1,2.6) {\scriptsize $1_\mathrm{out}$};
            \node at (4.075,2.6) {\scriptsize $n_\mathrm{in}$};
            \node at (6.715,2.6) {\scriptsize $n_\mathrm{out}$};
            
    \end{tikzpicture}
    \caption{Schematic representation of the most general causal sequential tester. Tested devices are shown in blue and tester elements in orange. Causal order proceeds from left to right.}
    \label{fig:sequential}
\end{figure}

The unambiguous discrimination task is carried out by quantum testers, which in the sequential scenario depicted in Fig.~\ref{fig:sequential}, are represented by operators satisfying 

\begin{subequations}
\begin{align}
    T_?+\sum_{|r|=k} T_r &= \mathbbm{1}_{n_\mathrm{out}}\otimes R^{(n)}\label{subeq:completeness} \\ 
    \mathrm{tr}_{i_\mathrm{in}}R^{(i)}&=\mathbbm{1}_{(i-1)_\mathrm{out}}\otimes R^{(i-1)}\quad i=2,\ldots,n\label{subeq:first-comb}\\
    \mathrm{tr}_{1_\mathrm{in}}R^{(1)}&=1\label{subeq:last-comb}\\
    T_i &\geq 0,\quad \forall i\label{subeq:positivity}
\end{align}
\end{subequations}
where $T_r$ denotes the part of the tester associated with identifying the anomaly at positions labeled by $r$ and $T_?$ associates with the inconclusive outcome. These conditions ensure complete positivity and causality of the (generalized) quantum tester~\cite[section IV]{chiribella_theoretical_2009}. In this setting, the average success probability is given by the generalized Born rule
\begin{equation}
P_{\mathrm{s}}=\int\mathrm{d}U\sum_{r}\frac{1}{N}\,\mathrm{tr}\big(T_r^T\tilde{F}_r\big)\,,
\label{eq:6-P_s-avg}
\end{equation}
where the superscript $^T$ denotes partial transposition in the computational basis and the operator $\tilde{F}_r$ denotes the Choi matrix of channel $\tilde{\mathcal{F}}_r$. The average is taken with respect to the Haar measure \cite{collins_integration_2006,mele_introduction_2024}, since the anomalous unitaries $U$ are unknown and thus are assumed to be uniformly distributed. Since the unitary $U$ defining the anomalies is unknown, we demand that the choice of the tester must be independent from it. This allows us to define the effective Choi matrices $F_r$ as
\begin{equation}
F_r =\int\mathrm{d}U \tilde{F}_r =\iddk^{\otimes(n-k)}_{\bar{r}} \otimes C^{(k)}_{r}\,,
\end{equation}
where $\iddk$ denotes the Choi matrix of the identity channel. For simplicity, single indices encompass both the ``in" and ``out" subspaces where operators act when no confusion arises. The matrix $C^{(k)}_r$ captures the action of the anomalous unitaries and is given by
\begin{align}\label{eq: General Choi}
C^{(k)}_{r} = \int (\mathbbm{1}_{r_\mathrm{in}}^{\otimes k} \otimes U_{r_\mathrm{out}}^{\otimes k})\, \iddk^{\otimes k}_r\, (\mathbbm{1}_{r_\mathrm{in}}^{\otimes k} \otimes U_{r_\mathrm{out}}^{\otimes k})^\dagger \,\mathrm{d}U\,,
\end{align}
which admits a closed form when $k\leq d$
\begin{equation}
C^{(k)}_{r} = \sum_{\sigma, \pi \in S_k} \mathrm{Wg}(\sigma \pi^{-1}, d)\, U_{r_\mathrm{in}}(\sigma) \otimes U_{r_\mathrm{out}}^*(\pi)\,,
\label{eq:6-weingarten}
\end{equation}
where $S_k$ is the symmetric group on $k$ elements, ${U}(\sigma)$ is the unitary representation of the permutation $\sigma\in S_k$ in a $k$-partite Hilbert space of local dimension $d$, and $\mathrm{Wg}(\sigma,d)$ is the Weingarten function~\cite{weingarten_asymptotic_1978,collins_weingarten_2022}.

\section{Optimal zero-error protocol}
The observer’s objective is the unambiguous identification of anomalous devices. 
In general, implementation of a quantum tester requires a sequential (or adaptive) strategy, in which a bipartite system plus a memory state is 
prepared and the action of the inserted quantum channels on a system is interleaved with bipartite quantum channels, which prepare input to the next inserted channel and also alter the memory system. As a final step system and memory are collectively measured.
The SDP formulation for the zero-error discrimination of quantum channels via 
quantum testers reads

\begin{subequations}
    \begin{align}
    &\hspace{-0.85cm}P_\mathrm{s}={\frac{1}{N}}\max_{\{T_r\}}\displaystyle\sum_{|r|=k}\mathrm{tr}(T_r^{T}F_r)\label{subeq:6-maximization}\\
    \mathrm{s.t.}\quad&\mathrm{tr}(T_r^{T}F_s)=0 \quad\forall r\neq s\label{subeq:6-unambiguouscondition}\\
    &T_r\geq0\label{subeq:6-positivecondition}\\
    &T_?+\displaystyle\sum_{{|r|=k}}T_r=\mathbbm{1}_{n_\mathrm{out}}\otimes R^{(n)}\label{subeq:6-firstcombcondition}\\
    &\mathrm{tr}_{i_\mathrm{in}}(R^{(i)})=\mathbbm{1}_{(i-1)_\mathrm{out}}\otimes R^{(i-1)}\quad i=2,\ldots,n\label{subeq:6-secondcombcondition}\\
    &\mathrm{tr}_{1_\mathrm{in}}(R^{(1)})=1\,.\label{subeq:6-normalizationcondition}
    \end{align}
    \label{eq:6-conditions}
\end{subequations}
The constraint in Eq.~\eqref{subeq:6-unambiguouscondition} is the zero-error condition for channel identification and the rest arise from the causal relations of quantum testers [c.f. Eqs.~(\ref{subeq:completeness}--\ref{subeq:positivity})]. 

\subsection{Single anomaly}
\label{sec:6-one-anomaly}

We begin by considering the case of a single anomaly, i.e., $k=1$, which implies $N=n$. In this setting, the Choi matrices $\{F_r\}$ exhibit a very convenient symmetry: each $F_r$ is invariant under the action of local unitaries $U\otimes U^*$ acting on the in–out pair of every bipartite system $j\in \{1,\ldots,n\}$. Explicitly, they satisfy
 
\begin{align}
    F_r=\hspace{-0.2cm}\displaystyle\bigintss \hspace{-0.1cm}\left(\bigotimes_{j=1}^N(U_{j}\otimes U_{j}^*)\right) F_r \left(\bigotimes_{j=1}^N(U_{j}\otimes U_{j}^*)^\dagger\right)\prod_{j=1}^N\mathrm{d}U_j\,.
    \label{eq:6-invariance}
\end{align}
This symmetry follows from the fact that $\iddk$ is invariant under the isotropic twirling $U\otimes U^*$, and the Haar averaging in Eqs. (\ref{eq: General Choi}--\ref{eq:6-weingarten}) for $k=1$ produces a Choi matrix proportional to the identity operator, which itself is invariant under any unitary transformation.

Without loss of generality, we can restrict our search for optimal testers to those that respect the same symmetry. One can simply define a symmetrized tester, i.e., a Haar averaged tester with respect to the above symmetry and show that the averaging will not affect Eqs.~(\ref{subeq:6-maximization}--\ref{subeq:6-normalizationcondition}).
We define the projectors $\Pi_0 = \frac{1}{d}\iddk$ and $\Pi_1 = \mathbbm{1} \otimes \mathbbm{1} - \Pi_0$, and consider operators of the form
\begin{equation}
    E_s=\bigotimes_{j=1}^n\big(\mathit{1}_{\bar{s}}(j)\,\Pi_0+\mathit{1}_{{s}}(j)\Pi_1\big)\,,
\end{equation}
where $\mathit{1}_s(j)$ is the indicator function 
\begin{equation}
    \mathit{1}_s(j)=\begin{cases}
        1\quad\mathrm{if}\hspace{0.2cm}j\in s\,,\\
        0\quad\mathrm{otherwise}\,.
    \end{cases}
\end{equation}
That is, the operator $E_s$ is a tensor product with $\Pi_1$ in the parties that belong to the set $s$ and $\Pi_0$ in the remainder of the systems. 
Using these operators, the symmetrized testers can be expressed as
\begin{equation}
T_r = \sum_{s \subseteq \{1,\ldots,n\}} a^r_s\, E_s\,,
\end{equation}
with $a_s^r$ a set of non-negative coefficients. The same basis also allows us to expand the Choi matrices $\{F_r\}_r$ as
\begin{equation}
F_r = d^{\hspace{0.04cm}n-2}(E_\emptyset + E_r)\,,
\end{equation}
where $\emptyset$ denotes the empty set.

Let us inspect in this basis implications of Eq. (\ref{subeq:6-unambiguouscondition}).
For all $r,s=1,\ldots,n$ and $r\neq s$, we must have $0=\mathrm{tr}(T_r^{T}F_s)=a^r_\emptyset + d^{n-2}(d^2-1)a^r_s$. Since coefficient $a^r_s$ are non-negative, this implies that for all $r\neq s$, $a^r_\emptyset=0$, $a^r_s=0$.
The success probability in Eq.~\eqref{subeq:6-maximization} then reduces to the simple expression
\begin{equation}
    P_\mathrm{s}=\frac{1}{n}d^{\,n}\frac{d^{\hspace{0.04cm}2}-1}{d^{\hspace{0.04cm}2}}\sum _{|r|=1}a_r^r\,,
\end{equation}

Defining the coefficients $\alpha_s=a_s^?+\sum_{|r|=1}a_s^r$, 
the 
tester normalization condition (\ref{subeq:6-firstcombcondition}) becomes
\begin{equation}
    T_?+\sum_{|r|=1}T_r=\mathbbm{1}_{n_\mathrm{out}}\otimes R^{(n)}=\sum_{s\subseteq\{1,\ldots,n\}}\alpha_s\,E_s\,,
    \label{eq:6-completeness-sym}
\end{equation}
Since linear combinations of projectors $\Pi_0$ and $\Pi_1$ can  create a factorized operator only if both are multiplied by the same constant, Eq. (\ref{eq:6-completeness-sym}) implies that 
\begin{equation}
\forall s \subseteq \{1,\ldots,n-1\}\;\; \alpha_{s \cup \{n\}} = \alpha_s\,,
\end{equation}
and consequently
\begin{equation}
    R^{(n)}=\mathbbm{1}_{{n_\mathrm{in}}}\otimes \sum_{s\subseteq\{1,\ldots,n-1\}}\alpha_s\,\mathrm{tr}_n(E_s)\,,
\end{equation}

On the other hand, the constraint $\mathrm{tr}_{n_\mathrm{in}} R^{(n)} = \mathbbm{1}_{n_\mathrm{out}} \otimes R^{(n-1)}$ implies

\begin{equation}
\forall s \subseteq \{1,\ldots,n-2\}\;\; \alpha_{s \cup \{n-1\}} = \alpha_s\,,
\end{equation}
and
\begin{equation}
    R^{(n-1)}=\mathbbm{1}_{{{n-1}_\mathrm{in}}}\otimes \sum_{s\subseteq\{1,\ldots,n-2\}} d\; \alpha_s\,\mathrm{tr}_{n,n-1}(E_s)\,,
\end{equation}

Iterating the same reasoning for all constraints in Eqs.~(\ref{subeq:6-secondcombcondition}--\ref{subeq:6-normalizationcondition}) we obtain that all $\alpha_s$ must be equal, i.e., $\forall s\; \alpha_s=\alpha$.
In particular, from Eqs.~(\ref{subeq:6-secondcombcondition}--\ref{subeq:6-normalizationcondition}) for $i=2$ we get $\alpha_\emptyset=\alpha_{\{1\}}=\alpha=d^{-n}$.
This implies that Eq. (\ref{eq:6-completeness-sym}) equals $\frac{1}{d^n}\otimes_{i=1}^n \mathbbm{1}_i$.

Since $a_r^r \leq \alpha = d^{-n}$, the optimal success probability is bounded by
\begin{equation}
    P_\mathrm{s}\leq1-\frac{1}{d^{\hspace{0.04cm}2}}\,.
\end{equation}
This bound is achievable by taking $T_r = d^{-n} E_r$ for each $|r|=1$, and $T_?=d^{-n}\mathbbm{1}^{\otimes 2n}-\sum_{|r|=1}T_r$. 

\begin{figure}[h!]
    \centering
    \begin{tikzpicture}[scale=0.5/2*1.5]
            \setlength{\unitlength}{0.1mm}
            \fill[yago3!70] (-1.25,0) rectangle (0.25,4);
            \draw[-] (-1.25,0) rectangle (0.25,4);
            \draw[-] (0.25,2+0.5) -- (2,2+0.5);
            \node at (0.6-1,2) {$\phi^+$};

            \node at (1.125,3) {\scriptsize {$1_\mathrm{in}$}};
            
            \fill[blue!25] (2,1+0.5) rectangle (4,3+0.5);
            \node at (3,2+0.5) {$\mathcal{D}_1$};
            \draw[-] (2,1+0.5) rectangle (4,3+0.5);

            \node at (4.875,3) {\scriptsize {$1_\mathrm{out}$}};
            
            \draw[-] (4,2+0.5) -- (5.75,2+0.5);
            \draw[-] (0.25,1) -- (5.75,1);
            \fill[yago3!70] (5.75,0) rectangle (7.25,4);
            \draw[-] (5.75,0) rectangle (7.25,4);
            \node at (6.5,2) {\scalebox{0.9}{$\mathcal{M}$}};

            \fill[yago3!70] (-1.25,0-5) rectangle (0.25,4-5);
            \draw[-] (-1.25,0-5) rectangle (0.25,4-5);
            \draw[-] (0.25,2+0.5-5) -- (2,2+0.5-5);
            \node at (0.6-1,2-5) {$\phi^+$};

            \node at (1.125,3-5) {\scriptsize {$2_\mathrm{in}$}};
            
            \fill[blue!25] (2,1+0.5-5) rectangle (4,3+0.5-5);
            \node at (3,2+0.5-5) {$\mathcal{D}_2$};
            \draw[-] (2,1+0.5-5) rectangle (4,3+0.5-5);

            \node at (4.875,3-5) {\scriptsize {$2_\mathrm{out}$}};
            
            \draw[-] (4,2+0.5-5) -- (5.75,2+0.5-5);
            \draw[-] (0.25,1-5) -- (5.75,1-5);
            \fill[yago3!70] (5.75,0-5) rectangle (7.25,4-5);
            \draw[-] (5.75,0-5) rectangle (7.25,4-5);
            \node at (6.5,2-5) {\scalebox{0.9}{$\mathcal{M}$}};

            \fill[yago3!70] (-1.25,0-11) rectangle (0.25,4-11);
            \draw[-] (-1.25,0-11) rectangle (0.25,4-11);
            \draw[-] (0.25,2+0.5-11) -- (2,2+0.5-11);
            \node at (0.6-1,2-11) {$\phi^+$};

            \node at (1.125,3-11) {\scriptsize {$n_\mathrm{in}$}};
            
            \fill[blue!25] (2,1+0.5-11) rectangle (4,3+0.5-11);
            \node at (3,2+0.5-11) {$\mathcal{D}_n$};
            \draw[-] (2,1+0.5-11) rectangle (4,3+0.5-11);

            \node at (4.875,3-11) {\scriptsize {$n_\mathrm{out}$}};
            
            \draw[-] (4,2+0.5-11) -- (5.75,2+0.5-11);
            \draw[-] (0.25,1-11) -- (5.75,1-11);
            \fill[yago3!70] (5.75,0-11) rectangle (7.25,4-11);
            \draw[-] (5.75,0-11) rectangle (7.25,4-11);
            \node at (6.5,2-11) {\scalebox{0.9}{$\mathcal{M}$}};
            
            \node at (3,-5.5) {\large \vdots};
        \end{tikzpicture}
    \caption{Diagrammatic representation of the optimal zero-error anomaly detection protocol. Independent measurements of each device show the local and parallel features of the scheme.}
    \label{fig:local-protocol}
\end{figure}
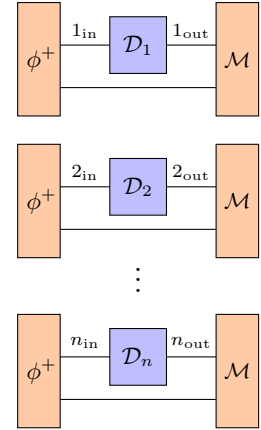

A single anomaly detection protocol achieving this bound proceeds as follows. One prepares product state $\ket{\Psi}=\ket{\phi^+}^{\otimes n}$, where $\ket{\phi^+}=\frac{1}{\sqrt{d}}\sum_{i}\ket{ii}$ denotes the maximally entangled state of two $d$-dimensional systems. The first subsystem of each copy of $\ket{\phi^+}$ is sent through one of the devices $\mathcal{D}_j$, while the second subsystem is left untouched. In each of these bipartite systems, a ``local" measurement 
is then performed to check whether the resulting state has projection onto $\ket{\phi^+}$ or not, corresponding to the application of the two-outcome measurement $\mathcal{M}=\{\Pi_0,\Pi_1\}$. If the outcome ``$1$" is obtained, the corresponding device is identified with certainty as anomalous; if the outcome is instead ``$0$", the remaining devices are tested in the same way until the anomaly is located. If none of the local measurements clicks ``$1$", the protocol output is inconclusive. This realization is depicted in Fig~\ref{fig:local-protocol}.

This protocol is local, in the sense that measurements act independently on each tested device 
and it can be applied in an online fashion, allowing the anomaly to be detected without the need to test all devices. 
Also, in the limit of large 
dimension $d$, the success probability approaches one. This is because the outputs of random unitaries tend to become more orthogonal as the Hilbert space dimension increases, making the anomalies easier to detect. Finally, it is noteworthy that we have considered the most general causal strategy for the protocol [c.f. Eqs.~(\ref{subeq:6-firstcombcondition}--\ref{subeq:6-normalizationcondition})]; however, in the end, it turns out that the optimal implementation corresponds to a parallel strategy. 
In general, it is not known in which general tasks involving unknown unitaries parallel implementations are sufficient and can be assumed beforehand. For instance, in~\cite{quintino_reversing_2019}, a parallel strategy suffices for the probabilistic inversion of a single use of an unknown unitary, while for multiple uses, adaptive strategies perform better. Similarly, in~\cite{sedlak_optimal_2019}, parallel implementations are shown to be optimal for the storage of $m$ unitaries. In contrast, in the context of unitary discrimination, ancillary systems are unnecessary when the unitaries are known~\cite{acin_statistical_2001,skotiniotis_identification_2024}. 
On the other hand, in the setting of 
zero-error single anomaly detection studied here, we proved that the optimal protocol is parallel and requires ancillary systems.

\subsection{Multiple anomalies}
\label{sec:6-two-anomaly}

We now turn to the more challenging scenario of multiple anomalies. In the single-anomaly scenario of the previous section, the symmetry of the Choi matrices allowed us to restrict the search for optimal testers to a small subspace. That symmetry was local and factorized across all parties, which greatly simplified both the analysis and the construction of optimal protocols for arbitrary local dimension $d$ and for the most general causal strategies.

This situation changes when considering, for example, two anomalies, $k=2$. Here, the Choi matrix of the two anomalous channels takes the form
\begin{multline}
    C^{(2)}=\dfrac{1}{d^2-1}\Big(\mathbbm{1}_\mathrm{in}\otimes\mathbbm{1_\mathrm{out}}-\frac{1}{d}\,\mathbbm{1}_\mathrm{in}\otimes\mathrm{P}_\mathrm{out}\\
    -\frac{1}{d}\,\mathrm{P}_\mathrm{in}\otimes\mathbbm{1}_\mathrm{out}+\mathrm{P}_\mathrm{in}\otimes\mathrm{P}_\mathrm{out}\Big),
\end{multline}
where $\mathrm{P}=(\iddk)_{12}^{T_1}$ denotes the representation of the SWAP operator: $\mathrm{P}\ket{i}\ket{j}=\ket{j}\ket{i}$, and  ``in" and ``out" labels account for both parties.

Unlike in the single-anomaly case, this matrix does not possess the same local symmetry on each bipartite subsystem. Instead, for an arbitrary number $k\geq2$ of anomalies, the symmetry of the hypotheses is given by the action of a global tensor product unitary, namely
\begin{align}
F_r=\hspace{-0.2cm}\displaystyle\bigintss \hspace{-0.1cm}\Big(U_\mathrm{in}\otimes U_\mathrm{out}^*\Big)^{\hspace{-0.075cm}\otimes n} F_r\, \Big(U_\mathrm{in}^\dagger\otimes U_\mathrm{out}^T\Big)^{\hspace{-0.075cm}\otimes n}\mathrm{d}U\,,
    \label{eq:6-multipleinvariance}
\end{align}
which does not factorize across $n$ tested devices. 
As a consequence, the space of symmetric operators compatible with this invariance is larger
and the simplifications available in the single-anomaly analysis no longer apply here.

The space of invariant operators under the transformation in Eq.~\eqref{eq:6-multipleinvariance} corresponds to the algebra of partially transposed permutations, $\mathcal{A}_{n,n}^d$~\cite{cox_blocks_2008,grinko_gelfand-tsetlin_2023,balanzo-juando_positive_2024,studzinski_irreducible_2025,grinko_mixed_2025}. This algebra plays a central role in our analysis, as it gives the structure for the operators commuting with the symmetry constraints given by Eq.~\eqref{eq:6-multipleinvariance}. This commuting relation is captured by the so-called mixed Schur–Weyl duality~\cite{turaev_operator_1990,koike_decomposition_1989,grinko_mixed_2025}. In contrast with the classical Schur-Weyl duality, which involves the symmetric group acting in duality with $\mathrm{SU}(d)$, here the algebra of partially transposed permutations is the commutant corresponding to $\mathrm{SU}(d)$ when considering the dual unitary action $U^*$ in some subspaces, adapted to the transposed structure. This duality provides a convenient basis for decomposing these invariant operators in a block-diagonal form, offering a convenient framework to exploit the symmetry in the analysis of the discrimination task (see Appendix~\ref{app:brauer} for details). 

However, due to this broader symmetry class, finding the optimal protocol for general $d$ and general testers becomes substantially more challenging. In what follows, we focus on the qubit case ($d=2$) of $k=2$ anomalies and restrict the optimization to parallel quantum testers, and follow the discussion with an extension to an arbitrary number of anomalies.

\subsubsection{Two anomalies}

When considering the scenario of $k=2$ anomalies, the symmetry constraints and feasibility conditions in the SDP take a simpler form, allowing for a fully analytical solution. More importantly, for $k \leq n/2$---a natural regime given that $k$ represents the number of anomalies---numerical checks for small values of $n$ ($n=4$ and $n=5$) show that parallel and sequential strategies yield the same optimal success probability. While a formal proof is still missing, the evidence for small cases suggests that, in this scenario, sequential strategies offer no advantage over parallel ones. 
This justifies focusing on the parallel setting. Under this assumption, we now show that, for \(d=2\) and \(k=2\), the local protocol introduced for the single-anomaly case remains optimal.

The proof proceeds in two steps. First, we verify that the proposed testers form a feasible solution to the primal SDP.
Then, we construct a suitable dual ansatz to demonstrate strong duality, which allows us to establish the optimality of the protocol. The explicit proof will be carried out for the case $n=4$, and we will use this result for proving optimality for arbitrary values of $n$.

The protocol is defined by testers of the form \mbox{$T_r=\frac{1}{d^n}E_r$}, which yields a success probability
\begin{equation}\label{eq:two-anomalies-Ps}
    P_\mathrm{s}=\frac{1}{d^2}\,\mathrm{tr}\hspace{-0.05cm}\left(C^{(2)}\Pi_1^{\otimes2}\right)=\frac{d^4-2d^2+2}{d^4}
    \, ,
\end{equation}
which, for $d=2$, reduces to
\begin{equation} 
    P_\mathrm{s}=\frac{5}{8}\, .
\end{equation}

As discussed in Section~\ref{sec:6-one-anomaly}, these testers satisfy all constraints in Eqs.~(\ref{subeq:6-maximization}--\ref{subeq:6-normalizationcondition}), as the parallel-comb setting is a special case of the adaptive implementation.
To prove optimality, we construct an ansatz for the dual problem whose value matches that of the primal.

The dual formulation for parallel testers reads 
\begin{subequations}
    \begin{align}
    &\hspace{-1.1cm}\min\limits_{y,\,Y,\,\{\nu_{sr}\}} \hspace{0.2cm} y\label{subeq:6-dual-minimization}\\
    \mathrm{s.t.}\quad &Y - \frac{1}{N} F_r^T+\sum_{s\neq r}\nu_{sr}F_s^T \geq 0\ \forall\ r\label{subeq:6-dual-unambiguouscondition}\\
    &y\mathbbm{1}_\mathrm{in}-\mathrm{tr}_{\mathrm{out}}Y\geq 0\label{subeq:6-dual-combcondition}\\
    &Y=Y^\dagger\label{subeq:6-dual-positivecondition}\\
    &y\in\mathbb{R}\label{subeq:6-dual-realvariablecondition}\\
    &\nu_{sr}\in\mathbb{R}\quad\forall\ r,s\label{subeq:6-dual-realunambiguouscondition}
    \end{align}
    \label{eq:6-dual-conditions}
\end{subequations}
The constraint in Eq.~\eqref{subeq:6-dual-combcondition} suggests that $Y$ should be chosen to saturate the inequality, with $\mathrm{tr}\,Y=y\,d^n$, and $y=5/8$. A natural starting point is
\begin{equation}
    Y=\frac{d^n}{N}\sum_{|r|=2}\sqrt{T_r}\,F_r\sqrt{T_r}\,.
\end{equation}
This choice satisfies Eqs.~(\ref{subeq:6-dual-combcondition}--\ref{subeq:6-dual-realvariablecondition}), but violates the constraint in Eq.~\eqref{subeq:6-dual-unambiguouscondition} for any real coefficients $\{\nu_{rs}\}$. This violation is known in the literature for the case of zero-error inference (see, e.g., \cite{diebra_quantum_2025}). The infimum of the dual problem is finite but not attainable---yet strong duality still holds, since feasible dual points can be constructed arbitrarily close to it. 
Moreover, note that $Y$ is invariant under bipartite system permutations; therefore, it is sufficient to consider a fixed $r$ in Eq.~\eqref{subeq:6-dual-unambiguouscondition} to check feasibility for all $r$. 

We consider the matrix $Y-\frac{1}{N}F_{r}+\sum_{s\neq r}\nu_{rs} F_s$, for a fixed set of anomalies $r$, which by construction lies in the algebra of partially transposed permutations. This property, conveniently, allows us to use mixed Schur-Weyl duality and rewrite the operator in a block diagonal form
\begin{equation}
    Y-\frac{1}{N}F_{r}+\sum_{s\neq r}\nu_{rs} F_s=\bigoplus_{\hat{\lambda}\in \hat{\mathcal{A}}_{4,4}^2}\mathbbm{1}_{\hat{\lambda}}\otimes \Xi^{\hat{\lambda}}_r\,,\label{eq:6-feasibility}
\end{equation}
where $\hat{\lambda}$ labels the irreducible representations (irreps for short) of the joint action of $\mathcal{A}_{4,4}^2$ and $\mathrm{SU}(2)$. These labels are usually identified with pairs of partitions $(\lambda_L,\lambda_R)$ that fulfill $\ell(\lambda_L)+\ell(\lambda_R)\leq 2$, depicted in their Young tableaux form in Fig.~\ref{fig:6-irreps-WBA} below (see Appendix~\ref{app:brauer} for a detailed discussion). 
However, as every element of the left-hand side of Eq.~\eqref{eq:6-feasibility} contains a double contraction, i.e., terms of the form $\iddk^{\otimes 2}$, only a restricted set of irreps contributes. In fact, non-vanishing components appear exclusively in partitions $\big((2),(2)\big)$, $\big((1),(1)\big)$, and $(\varnothing,\varnothing)$. This follows from the way contractions affect the partitions: each contraction effectively reduces the left Young diagram ($\lambda_L$) by one box, thereby restricting which irreps can survive~\cite{grinko_gelfand-tsetlin_2023}. As a consequence, all other irreps vanish and can be discarded. This observation is the key to the dimensionality reduction, although the operator formally lives in a Hilbert space of dimension $2^8$, its relevant information is confined to the direct sum of just three components in the space corresponding to the representation of the algebra of partially transposed permutations, $\Xi^{\hat{\lambda}}_r$, which makes the subsequent analysis tractable.
\begin{figure}[ht!]
    \centering
\begin{tikzpicture}[scale=0.6, every node/.style={font=\normalsize}]
  \node[draw=none, fill=none] (const) at (0,0) {{ \textcolor{black}{$\raisebox{0.05cm}{(}\ \scalebox{0.5}{\ydiagram{4}}\ ,\,\scalebox{0.5}{\ydiagram{4}}\ \raisebox{0.05cm}{)}$}}};
  \node[draw=none, fill=none] (const) at (6,0) {{ \textcolor{black}{$\raisebox{0.05cm}{(}\ \scalebox{0.5}{\ydiagram{3}}\ ,\,\scalebox{0.5}{\ydiagram{3}}\ \raisebox{0.05cm}{)}$}}};
  \node[draw=none, fill=none] (const) at (-2,-1.75) {{ \textcolor{black}{$\raisebox{0.05cm}{(}\ \scalebox{0.5}{\ydiagram{2}}\ ,\,\scalebox{0.5}{\ydiagram{2}}\ \raisebox{0.05cm}{)}$}}};
  \node[draw=none, fill=none] (const) at (3.25,-1.75) {{ \textcolor{black}{$\raisebox{0.05cm}{(}\ \scalebox{0.5}{\ydiagram{1}}\ ,\,\scalebox{0.5}{\ydiagram{1}}\ \raisebox{0.05cm}{)}$}}};
  \node[draw=none, fill=none] (const) at (8.6,-1.75) {{\textcolor{black}{\raisebox{0.05cm}{$(\varnothing\ ,\,\varnothing)$}}}};
\end{tikzpicture}
    \caption[Paris of Young diagrams associated with the irreps of $\mathrm{SU}(2)$ and $\mathcal{A}_{4,4}^2$.]{Pairs of Young diagrams associated with the irreps of the joint action of $\mathrm{SU}(2)$ and $\mathcal{A}_{4,4}^2$. The first row contains the spaces in which the elements of Eq.~\eqref{eq:6-feasibility} have vanishing representation, while the second row contains the spaces in which the representation is non-trivial.}
    \label{fig:6-irreps-WBA}
\end{figure}

Another important reduction is that, due to the symmetry of the constraint in Eq.~\eqref{eq:6-feasibility}, we can set $\nu_{rs}=\nu$ for every $s\neq r$. It is easy to see that for sufficiently large values of $\nu$, the operators $\Xi_{\,r}^{\left((2),(2)\right)}$ and $\Xi_{\,r}^{\left((1),(1)\right)}$ are positive semidefinite. The case of $\Xi_{\,r}^{\left(\varnothing,\varnothing\right)}$ is more complex, and turns out that all its eigenvalues are non-negative, except for one, which behaves as $-\frac{k^{(1)}}{\nu}+\mathcal{O}\left(\nu^{-2}\right)$, where $k^{(1)}$ is a constant of $\mathcal{O}(1)$.

This allows us to introduce the new dual variables
\begin{equation}
    Y_\epsilon=\dfrac{1}{N}\sum_{|r|=k}\sqrt{T_r}F_r\sqrt{T_r}+\frac{\epsilon}{d^8} \mathbbm{1}_\mathrm{out}\otimes\mathbbm{1}_\mathrm{in}\,,
\end{equation}
with $\epsilon\simeq\frac{k^{(1)}}{\nu}$, and $y_\epsilon=5/8+\epsilon$. This choice of dual variables constitutes a feasible instance of the SDP for sufficiently large values of $\nu$. Taking the limit \mbox{$\epsilon\to0^+$} ($\nu\to\infty$) yields a dual value arbitrarily close to $5/8$. Since the primal optimum cannot exceed the dual infimum,  this establishes strong duality and proves that, for qubits with 
$n=4$ and $k=2$, the optimal success probability is $P_\mathrm{s}=5/8$.

For an arbitrary number of devices $n$, the local protocol attains a constant success probability $P_\mathrm{s}=5/8$. To prove its optimality, we now show that this value is, in fact, an upper bound---which our protocol saturates. 

Let $P_\mathrm{s}^*(n)$ denote the optimal success probability for $n$ devices, and suppose that an external agent provides the additional information that the two anomalous devices are guaranteed to be among a fixed subset $\Tilde{\mathscr{D}}_4$ of four devices. We denote the optimal success probability for this scenario by $P^*_\mathrm{s}(n|\Tilde{\mathscr{D}}_4)$. Clearly, knowing $\Tilde{\mathscr{D}}_4$ can only help, so 
\begin{equation}
    P^*_\mathrm{s}(n)\leq P^*_\mathrm{s}(n|\Tilde{\mathscr{D}}_4)\,.
\end{equation}

Note that the hypotheses for this scenario are exactly those of $n=4$ tensored by $\iddk^{\otimes (n-4)}$, which represents the Choi matrices of the remaining devices. This extra factor carries no distinguishing information---as it is identical for all hypotheses---so both optimal success probabilities are equal: \mbox{$P^*_\mathrm{s}(n|\Tilde{\mathscr{D}}_4)=P^*_\mathrm{s}(4)$}. Combining these observations with the fact that the local protocol achieves $P_\mathrm{s}(n)=5/8$ for any $n$ gives
\begin{equation}
    \frac{5}{8}\leq P^*_\mathrm{s}(n)\leq\frac{5}{8}\,,
\end{equation}
which proves the optimality of the protocol for all $n$. 

\subsubsection{Extension to an arbitrary number of anomalies}
We can tackle the general scenario of arbitrary number of anomalous devices and local dimension using the same local parallel strategy. The performance of this strategy in the general scenario can be explicitly expressed in terms of the local dimension of the subsystems as (see Appendix~\ref{app:local-strategy-general})
\begin{multline}
    P_s(k,d)=\frac{1}{d^k}\mathrm{tr}\left(C^{(k)}\Pi_1^{\otimes k}\right)
    \\
    = \frac{1}{d^{2k}}\sum_{m=0}^k(-1)^m\binom{k}{m}f_{m,d}\,d^{2(k-m)}\,.
    \label{eq:general-ps}
\end{multline}
where the coefficients $f_{m,d}$ are defined by the unitary group integral
\begin{equation}
    f_{m,d} = \int |\mathrm{tr}(U)|^{2m}\,dU.
\end{equation}
These moment integrals form a sequence of non-negative integers that can be regarded as the number of permutations of $S_m$ with the longest increasing subsequence of length at most $d$, in the context of combinatorics. More generally, they can be computed from the dimensions of the irreps of $S_m$~\cite{rains_increasing_1998}. 

In certain regimes, these coefficients admit simple closed-form expressions. For instance, when $d\geq m$, we have $f_{m,d}=m!$, which recovers Eq.~(\ref{eq:two-anomalies-Ps}) for the two-anomaly case. Remarkably, for qubit systems ($d=2$), these coefficients reduce to the Catalan numbers $f_{m,2}=C_m:=\frac{1}{2m+1}\binom{2m+1}{m}$~\cite{abramowitz_handbook_1948}, and the success probability simplifies significantly to
\begin{equation}
    \label{k-anomalies-qubits}
    P_s(k,d=2)=\frac{1}{2^{2k}}\binom{2k+1}{k+1}\,.
\end{equation}

We have shown that the local parallel strategy yields a nonvanishing success probability for an arbitrary number of devices. Establishing its strict optimality in the most general setting, however, remains an open problem. Nevertheless, our numerical results suggest that this strategy cannot be outperformed, even when more general sequential designs are allowed. We therefore conjecture that the local parallel strategy is optimal in the general multi-anomaly setting.

Several observations support this conjecture: the established optimality for the case of single and double anomalies, numerical results for higher values of $k$, and the fact that the zero-error protocol for the analogous state preparation problem is also solved by a parallel local strategy \cite{llorens_quantum_2024}. As with the two-anomaly case, constructing a formal proof would require the analysis of the high-dimensional operators arising from mixed Schur-Weyl duality and their diagonalization. Completely resolving the dual SDP for the general case remains technically challenging. Nonetheless, the underlying symmetries of the problem and its algebraic structure strongly suggest that our proposed strategy is indeed optimal.

\section{Summary and conclusions}
 
We have studied the problem of unambiguously identifying faulty quantum devices that are intended to implement a given unitary evolution. Our analysis focused on an agnostic setting, where the observer has no prior information about the faulty action beyond the fact that it is unitary. This lack of knowledge was modeled by assuming that all unitary evolutions are equally likely, thereby we defined effective hypotheses via Haar integration. The resulting protocol is universal, as it is not tailored to any particular transformation, but works for all possible deviations.

We derived the optimal protocols for the cases of one and two faulty devices. For the case of a single anomaly, we proved that parallel strategies suffice to attain the optimal zero-error success probability. 
This success probability is independent of the total number of devices, $n$, and tends to unity in the large-$d$ limit. In the case of two anomalies, we 
focused on 
qubits and parallel strategies, motivated by numerical evidence suggesting that adaptive methods offer no advantage over parallel ones.
We established that the very same protocol introduced for the single-anomaly case remains optimal in this setting. An appealing feature of this protocol is its online applicability: each device is measured independently of the others, and the procedure can be halted as soon as the $k$ anomalies have been detected, without the need to test all devices.

From a technical perspective, our work provides explicit optimal protocols and success probabilities for the identification of multiple unknown anomalies in a sequence of quantum channels. These results are significant, as analytic solutions to multi-process discrimination problems are scarce in the literature. Here we obtained the optimal solution to a universal task using representation-theoretic tools, specifically the algebra of partially transposed permutations and mixed Schur–Weyl duality. In doing so, we lift techniques from universal state discrimination~\cite{sentis_programmable_2013,llorens_quantum_2024,llorens_quantum_2025} to the more general setting of quantum channels.

Our findings suggest several promising directions for future research. Throughout this work we assumed that devices act on separate, independent systems. In realistic scenarios, such as quantum computers or more general quantum networks, a single quantum system may instead undergo a sequence of successive operations. Extending the present framework to the identification of faulty devices acting sequentially on the same system represents a natural next step. We obtained the analytic expression for the success probability of the extended parallel strategy---originally optimized for one and two anomalies---for an arbitrary number of anomalies. There are strong indications that Eq.~\eqref{eq:general-ps} is optimal in this general setting, and an analytical proof of this optimality would substantially broaden the scope of our results. In a similar vein, the protocols developed here are single-shot strategies; yet, in practice, a device may be available for repeated use. Allowing for multiple uses of the same device may improve the probability of correct identification. Characterizing the performance limits of such protocols, as well as the trade-off between resources (such as the number of input probes or device uses) and achievable performance, opens new avenues for future investigation.

\section{Acknowledgments}

This work has been financially supported by Ministerio de Ciencia e Innovación of the Spanish Government with funding from European Union
Next GenerationEU (PRTR-C17.l1) and by the Generalitat of Catalunya, by the Ministry of Economic Affairs and Digital Transformation of the Spanish Government through the QUANTUMENIA project: Quantum Spain, by the European Union through the Recovery, Transformation and Resilience Plan - NextGenerationEU within the framework of the “Digital Spain 2026 Agenda”, and by the Agencia Estatal de Investigación MCIN/AEI/10.13039/501100011033 with
project  
PID2022-
141283NB-I00 with the support of FEDER funds. S.L. was funded in whole or in part by the National Science Centre, Poland 2024/54/E/ST2/00451.
For the purpose of Open Access,
the author has applied a CC-BY public copyright licence to any Author Accepted Manuscript (AAM) version arising from this submission and also acknowledges the mobility program mobiQUTE. A.D. acknowledges funding from MCIN (FPU23/02763). M.S. also acknowledges 
support by the Slovak Research and Development Agency through the project APVV-22-0570 (DeQHOST), by project VEGA 2/0164/25 (Quantum Structures) and by project Cost CA22113. 
M.S. was partly funded by the EU NextGenerationEU through the Recovery and Resilience Plan for Slovakia under the project No. 09I03-03-V04-00777.

\bibliography{bibliography}

@article{llorens_quantum_2024,
	title = {Quantum multi-anomaly detection},
	volume = {8},
	issn = {2521-327X},
	url = {http://arxiv.org/abs/2312.13020},
	doi = {10.22331/q-2024-08-28-1452},
	language = {en},
	urldate = {2024-11-19},
	journal = {Quantum},
	author = {Llorens, Santiago and Sentís, Gael and Munoz-Tapia, Ramon},
	month = aug,
	year = {2024},
	pages = {1452},
}

@article{skotiniotis_identification_2024,
	title = {Identification of malfunctioning quantum devices},
	volume = {6},
	issn = {2643-1564},
	url = {https://link.aps.org/doi/10.1103/PhysRevResearch.6.033329},
	doi = {10.1103/PhysRevResearch.6.033329},
	language = {en},
	number = {3},
	urldate = {2024-11-19},
	journal = {Physical Review Research},
	author = {Skotiniotis, Michalis and Llorens, Santiago and Hotz, Ronja and Calsamiglia, John and Muñoz-Tapia, Ramon},
	month = sep,
	year = {2024},
	pages = {033329},
}

@article{chiribella_theoretical_2009,
	title = {Theoretical framework for quantum networks},
	volume = {80},
	copyright = {http://link.aps.org/licenses/aps-default-license},
	issn = {1050-2947, 1094-1622},
	url = {https://link.aps.org/doi/10.1103/PhysRevA.80.022339},
	doi = {10.1103/PhysRevA.80.022339},
	language = {en},
	number = {2},
	urldate = {2025-05-25},
	journal = {Physical Review A},
	author = {Chiribella, Giulio and D’Ariano, Giacomo Mauro and Perinotti, Paolo},
	month = aug,
	year = {2009},
	pages = {022339},
}

@article{diebra_quantum_2025,
	title={Quantum state exclusion for group-generated ensembles of pure states},
  author={Diebra, Arnau and Llorens, Santiago and Bagan, Emili and Sent{\'\i}s, Gael and Mu{\~n}oz-Tapia, Ramon},
  journal={Physical Review Research},
  volume={8},
  number={1},
  pages={L012001},
  year={2026},
  publisher={APS},
  doi={https://doi.org/10.1103/2k5d-bprn},
  url={https://doi.org/10.1103/2k5d-bprn}
}

@article{llorens_quantum_2025,
	title = {Quantum {Edge} {Detection}},
	volume = {9},
	issn = {2521-327X},
	url = {http://arxiv.org/abs/2405.11373},
	doi = {10.22331/q-2025-04-03-1687},
	language = {en},
	urldate = {2025-05-26},
	journal = {Quantum},
	author = {Llorens, Santiago and González, Walther and Sentís, Gael and Calsamiglia, John and Muñoz-Tapia, Ramon and Bagan, Emili},
	month = apr,
	year = {2025},
	pages = {1687},
}

@article{acin_statistical_2001,
    title = {Statistical {Distinguishability} between {Unitary} {Operations}},
    volume = {87},
    url = {https://link.aps.org/doi/10.1103/PhysRevLett.87.177901},
    doi = {10.1103/PhysRevLett.87.177901},
    number = {17},
    urldate = {2025-06-05},
    journal = {Physical Review Letters},
    author = {Acín, A.},
    month = oct,
    year = {2001},
    pages = {177901},
}

@article{collins_integration_2006,
    title = {Integration with {Respect} to the {Haar} {Measure} on {Unitary}, {Orthogonal} and {Symplectic} {Group}},
    volume = {264},
    issn = {1432-0916},
    url = {https://doi.org/10.1007/s00220-006-1554-3},
    doi = {10.1007/s00220-006-1554-3},
    language = {en},
    number = {3},
    urldate = {2025-07-08},
    journal = {Communications in Mathematical Physics},
    author = {Collins, Benoît and Śniady, Piotr},
    month = jun,
    year = {2006},
    pages = {773--795},
}

@article{mele_introduction_2024,
    title = {Introduction to {Haar} {Measure} {Tools} in {Quantum} {Information}: {A} {Beginner}'s {Tutorial}},
    volume = {8},
    shorttitle = {Introduction to {Haar} {Measure} {Tools} in {Quantum} {Information}},
    url = {https://quantum-journal.org/papers/q-2024-05-08-1340/},
    doi = {10.22331/q-2024-05-08-1340},
    language = {en-GB},
    urldate = {2025-07-08},
    journal = {Quantum},
    author = {Mele, Antonio Anna},
    month = may,
    year = {2024},
    pages = {1340},
}

@article{cox_blocks_2008,
    title = {On the blocks of the walled {Brauer} algebra},
    volume = {320},
    copyright = {https://www.elsevier.com/tdm/userlicense/1.0/},
    issn = {00218693},
    url = {https://linkinghub.elsevier.com/retrieve/pii/S0021869308000525},
    doi = {10.1016/j.jalgebra.2008.01.026},
    language = {en},
    number = {1},
    urldate = {2025-07-08},
    journal = {Journal of Algebra},
    author = {Cox, Anton and De Visscher, Maud and Doty, Stephen and Martin, Paul},
    month = jul,
    year = {2008},
    pages = {169--212},
}

@article{balanzo-juando_positive_2024,
    title = {Positive maps from the walled {Brauer} algebra},
    volume = {57},
    issn = {1751-8121},
    url = {https://dx.doi.org/10.1088/1751-8121/ad2b86},
    doi = {10.1088/1751-8121/ad2b86},
    language = {en},
    number = {11},
    urldate = {2025-07-08},
    journal = {Journal of Physics A: Mathematical and Theoretical},
    author = {Balanzó-Juandó, Maria and Studziński, Michał and Huber, Felix},
    month = mar,
    year = {2024},
    pages = {115202},
}

@article{studzinski_irreducible_2025,
      title={Group-Adapted Irreducible Matrix Units for the Walled Brauer Algebra}, 
      author={Michał Studziński and Tomasz Młynik and Marek Mozrzymas and Michał Horodecki and Dmitry Grinko},
      year={2026},
      journal={arXiv:2501.13067},
      url={https://arxiv.org/abs/2501.13067}, 
}

@article{grinko_gelfand-tsetlin_2023,
    title = {Gelfand-{Tsetlin} basis for partially transposed permutations, with applications to quantum information},
    url = {http://arxiv.org/abs/2310.02252},
    urldate = {2025-07-08},
    journal = {arXiv:2310.02252},
    author = {Grinko, Dmitry and Burchardt, Adam and Ozols, Maris},
    month = oct,
    year = {2023},
}

@article{sentis_programmable_2013,
    title = {Programmable discrimination with an error margin},
    volume = {88},
    url = {https://link.aps.org/doi/10.1103/PhysRevA.88.052304},
    doi = {10.1103/PhysRevA.88.052304},
    number = {5},
    urldate = {2025-07-22},
    journal = {Physical Review A},
    author = {Sentís, G. and Bagan, E. and Calsamiglia, J. and Muñoz-Tapia, R.},
    month = nov,
    year = {2013},
    pages = {052304},
}

@phdthesis{grinko_mixed_2025,
    title = {Mixed {Schur}–{Weyl} duality in quantum information},
    url = {https://core.ac.uk/download/pdf/645691932.pdf},
    urldate = {2025-07-23},
    publisher = {Institute for Logic, Language and Computation},
    author = {Grinko, Dmitry},
    school = {Institute for Logic, Language and Computation},
    year = {2025},
}

@article{koike_decomposition_1989,
    title = {On the decomposition of tensor products of the representations of the classical groups: by means of the universal characters},
    volume = {74},
    number = {1},
    journal = {Advances in Mathematics},
    author = {Koike, Kazuhiko},
    year = {1989},
    pages = {57--86},
    url = {https://doi.org/10.1016/0001-8708(89)90004-2},
}

@article{turaev_operator_1990,
    title = {Operator invariants of tangles, and {R}-matrices},
    volume = {35},
    number = {2},
    journal = {Mathematics of the USSR-Izvestiya},
    author = {Turaev, Vladimir G},
    year = {1990},
    pages = {411},
    url = {https://doi.org/10.1070/IM1990v035n02ABEH000711},
}

@book{abramowitz_handbook_1948,
    title = {Handbook of mathematical functions with formulas, graphs, and mathematical tables},
    publisher = {US Government printing office},
    author = {Abramowitz, Milton and Stegun, Irene A},
    year = {1968},
}

@book{tartakovsky_sequential_2014,
    title = {Sequential analysis: {Hypothesis} testing and changepoint detection},
    publisher = {CRC press},
    author = {Tartakovsky, Alexander and Nikiforov, Igor and Basseville, Michele},
    year = {2014},
}

@article{thottan_anomaly_2003,
    title = {Anomaly detection in {IP} networks},
    volume = {51},
    number = {8},
    journal = {IEEE Transactions on signal processing},
    author = {Thottan, Marina and Ji, Chuanyi},
    year = {2003},
    pages = {2191--2204},
    url = {https://doi.org/10.1109/TSP.2003.814797},
}

@inproceedings{siris_application_2004,
    title = {Application of anomaly detection algorithms for detecting {SYN} flooding attacks},
    volume = {4},
    booktitle = {{IEEE} {Global} {Telecommunications} {Conference}, 2004. {GLOBECOM}'04.},
    publisher = {IEEE},
    author = {Siris, Vasilios A and Papagalou, Fotini},
    year = {2004},
    pages = {2050--2054},
    url = {https://doi.org/10.1109/GLOCOM.2004.1378372},
}

@article{ahmed_survey_2016,
    title = {A survey of anomaly detection techniques in financial domain},
    volume = {55},
    journal = {Future Generation Computer Systems},
    author = {Ahmed, Mohiuddin and Mahmood, Abdun Naser and Islam, Md Rafiqul},
    year = {2016},
    pages = {278--288},
    url = {https://doi.org/10.1016/j.future.2015.01.001},
}

@article{zipfel_anomaly_2023,
    title = {Anomaly detection for industrial quality assurance: {A} comparative evaluation of unsupervised deep learning models},
    volume = {177},
    journal = {Computers \& Industrial Engineering},
    author = {Zipfel, Justus and Verworner, Felix and Fischer, Marco and Wieland, Uwe and Kraus, Mathias and Zschech, Patrick},
    year = {2023},
    pages = {109045},
    url = {https://doi.org/10.1016/j.cie.2023.109045},
}

@article{weingarten_asymptotic_1978,
    title = {Asymptotic behavior of group integrals in the limit of infinite rank},
    volume = {19},
    number = {5},
    journal = {Journal of Mathematical Physisc},
    author = {Weingarten, Don},
    year = {1978},
    url = {https://doi.org/10.1063/1.523807},
}

@article{collins_weingarten_2022,
    title = {The weingarten calculus},
    volume = {69},
    number = {05},
    journal = {Notices of the American Mathematical Society},
    author = {Collins, Benoit and Matsumoto, Sho and Novak, Jonathan},
    year = {2022},
    pages = {1},
    url = {https://doi.org/10.1090/noti2474},
}

@article{quintino_reversing_2019,
    title = {Reversing unknown quantum transformations: {Universal} quantum circuit for inverting general unitary operations},
    volume = {123},
    number = {21},
    journal = {Physical review letters},
    author = {Quintino, Marco Túlio and Dong, Qingxiuxiong and Shimbo, Atsushi and Soeda, Akihito and Murao, Mio},
    year = {2019},
    pages = {210502},
    url = {https://doi.org/10.1103/PhysRevLett.123.210502},
}

@article{sedlak_optimal_2019,
    title = {Optimal probabilistic storage and retrieval of unitary channels},
    volume = {122},
    number = {17},
    journal = {Physical Review Letters},
    author = {Sedlák, Michal and Bisio, Alessandro and Ziman, Mário},
    year = {2019},
    pages = {170502},
    url = {https://doi.org/10.1103/PhysRevLett.122.170502},
}

@article{choi_completely_1975,
    title = {Completely positive linear maps on complex matrices},
    volume = {10},
    number = {3},
    journal = {Linear algebra and its applications},
    author = {Choi, Man-Duen},
    year = {1975},
    pages = {285--290},
    doi={https://doi.org/10.1016/0024-3795(75)90075-0},
    url={https://doi.org/10.1016/0024-3795(75)90075-0}
}

@article{jamiolkowski_linear_1972,
    title = {Linear transformations which preserve trace and positive semidefiniteness of operators},
    volume = {3},
    number = {4},
    journal = {Reports on mathematical physics},
    author = {Jamiołkowski, Andrzej},
    year = {1972},
    pages = {275--278},
    doi={https://doi.org/10.1016/0034-4877(72)90011-0},
    url={https://doi.org/10.1016/0034-4877(72)90011-0}
}

@article{rains_increasing_1998,
    title = {Increasing subsequences and the classical groups},
    journal = {the electronic journal of combinatorics},
    author = {Rains, Eric M},
    year = {1998},
    pages = {R12--R12},
    doi={https://doi.org/10.37236/1350},
    url={https://doi.org/10.37236/1350}
}

@article{arute_quantum_2019,
    title = {Quantum supremacy using a programmable superconducting processor},
    volume = {574},
    copyright = {2019 The Author(s), under exclusive licence to Springer Nature Limited},
    issn = {1476-4687},
    url = {https://www.nature.com/articles/s41586-019-1666-5},
    doi = {10.1038/s41586-019-1666-5},
    language = {en},
    number = {7779},
    urldate = {2026-04-03},
    journal = {Nature},
    publisher = {Nature Publishing Group},
    author = {Arute, Frank and others},
    month = oct,
    year = {2019},
    pages = {505--510},
}

@article{nigg_quantum_2014,
    title = {Quantum computations on a topologically encoded qubit},
    volume = {345},
    url = {https://www.science.org/doi/10.1126/science.1253742},
    doi = {10.1126/science.1253742},
    number = {6194},
    urldate = {2026-04-03},
    journal = {Science},
    publisher = {American Association for the Advancement of Science},
    author = {Nigg, D. and Müller, M. and Martinez, E. A. and Schindler, P. and Hennrich, M. and Monz, T. and Martin-Delgado, M. A. and Blatt, R.},
    month = jul,
    year = {2014},
    pages = {302--305},
}

@article{corcoles_demonstration_2015,
    title = {Demonstration of a quantum error detection code using a square lattice of four superconducting qubits},
    volume = {6},
    issn = {2041-1723},
    url = {https://www.nature.com/articles/ncomms7979},
    doi = {10.1038/ncomms7979},
    language = {en},
    number = {1},
    urldate = {2026-04-03},
    journal = {Nature Communications},
    author = {Córcoles, A.D. and Magesan, Easwar and Srinivasan, Srikanth J. and Cross, Andrew W. and Steffen, M. and Gambetta, Jay M. and Chow, Jerry M.},
    month = apr,
    year = {2015},
    pages = {6979},
}

@article{liao_satellite--ground_2017,
    title = {Satellite-to-ground quantum key distribution},
    volume = {549},
    number = {7670},
    journal = {Nature},
    publisher = {Nature Publishing Group UK London},
    author = {Liao, Sheng-Kai and Cai, Wen-Qi and Liu, Wei-Yue and Zhang, Liang and Li, Yang and Ren, Ji-Gang and Yin, Juan and Shen, Qi and Cao, Yuan and Li, Zheng-Ping and {others}},
    year = {2017},
    pages = {43--47},
    doi ={https://doi.org/10.1038/nature23655},
    url={https://doi.org/10.1038/nature23655}
}

@article{khatri_principles_2020,
    title = {Principles of quantum communication theory: {A} modern approach},
    journal = {arXiv:2011.04672},
    author = {Khatri, Sumeet and Wilde, Mark M},
    year = {2020},
    url={https://arxiv.org/abs/2011.04672}
}

@article{degen_quantum_2017,
    title = {Quantum sensing},
    volume = {89},
    url = {https://link.aps.org/doi/10.1103/RevModPhys.89.035002},
    doi = {10.1103/RevModPhys.89.035002},
    number = {3},
    urldate = {2024-05-15},
    journal = {Reviews of Modern Physics},
    publisher = {American Physical Society},
    author = {Degen, C. L. and Reinhard, F. and Cappellaro, P.},
    month = jul,
    year = {2017},
    pages = {035002},
}

@article{reiter_dissipative_2017,
    title = {Dissipative quantum error correction and application to quantum sensing with trapped ions},
    volume = {8},
    number = {1},
    journal = {Nature communications},
    publisher = {Nature Publishing Group UK London},
    author = {Reiter, Florentin and Sørensen, Anders Søndberg and Zoller, Peter and Muschik, Christine A},
    year = {2017},
    pages = {1822},
    doi={https://doi.org/10.1038/s41467-017-01895-5},
    url={https://doi.org/10.1038/s41467-017-01895-5}
}

@misc{grinko_walledbrauer-gtbasis_2023,
    title = {walledbrauer-gtbasis},
    publisher = {GitHub},
    author = {Grinko, Dmitry and Burchardt, Adam and Ozols, Maris},
    year = {2023},
    note = {\url{https://github.com/dgrinko/walledbrauer-gtbasis} },
}

@misc{llorens_Exact-identification-unknown-unitary,
    title = {Exact-identification-unknown-unitary},
    publisher = {GitHub},
    author = {Llorens, Santiago and Diebra, Arnau and Sedlák, Michal and Muñoz-Tapia, Ramon},
    year = {2026},
    note = {\url{https://github.com/SantiagoLlorens/Exact-identification-unknown-unitary} },
}
\clearpage
\newpage

\onecolumngrid
\appendix

\section{Algebra of partially transposed permutations}\label{app:brauer}

In this appendix, we provide a concise introduction to the representation-theoretic framework underlying the symmetry analysis of the main text. We describe the algebra of partially transposed permutations and the walled Brauer algebra (WBA), which arise from the invariance of the hypotheses in the anomalous unitary detection problem.

\subsection*{Algebraic structure}
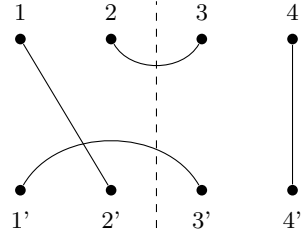
\begin{figure}[ht!]
    \centering
\begin{tikzpicture}[scale=1,
  every node/.style={circle, fill=black, minimum size=4pt, inner sep=0pt}]

  \def\sep{1.2} 
  \def\vertsep{2}  

  \foreach \i in {1,...,4} {
    \node[label={[label distance=4pt]above:{\i}}] (t\i) at (\i*\sep, \vertsep) {};
  }
  \foreach \i in {1,...,4} {
    \node[label={[label distance=4pt]below:{\i'}}] (b\i) at (\i*\sep, 0) {};
  }
  \draw[dashed] (2.5*\sep, -0.5) -- (2.5*\sep, \vertsep+0.5);

  \draw (t1) -- (b2);
  \draw (t2) to[out=-60, in=-120] (t3);
  \draw (b1) to[out=-300, in=-240] (b3);
  \draw (t4) -- (b4);

\end{tikzpicture}
    \caption{An example of an element $A$ of the walled Brauer algebra $\mathfrak{B}_{2,2}^d$, acting on $V^{\otimes 2}\otimes (V^*)^{\otimes 2}$. The dashed line separates the usual vector spaces from the dual vector spaces.}
    \label{fig:wba-element}
\end{figure}

The algebra of partially transposed permutations, $\mathcal{A}_{n,m}^d$, is the matrix representation of the WBA, $B_{n,m}(d)$, a certain subalgebra of the Brauer algebra $B_{n+m}(d)$. The elements of the WBA are intuitively represented by diagrams consisting of two horizontal rows---consisting on $n+m$ dots on the top row and $n+m$ dots on the bottom---separated by a vertical ``wall" that differentiates the first $n$ dots on the left from the remaining $m$ dots on the right. In these diagrams, every dot must be connected pairwise to exactly one other dot by a continuous line, subject to the following rules: a line connecting two dots within the same row must cross the wall, whereas a line connecting a top dot to a bottom dot is strictly forbidden from crossing the wall. An illustrative example of a valid WBA element is shown in Fig.~\ref{fig:wba-element}.

The algebra is generated by three types of elements acting on the tensor product space $(\mathbb{C}^d)^{\otimes n}\otimes (\mathbb{C}^d)^{\otimes m}$:
\begin{itemize}
    \item Left transpositions: $s_i$, with $i=1,\ldots,n-1$, which correspond to swapping the $i$-th subsystem with the $(i+1)$-th subsystem within the first $n$ subsystems, i.e., left to the wall.
    
    \item Contraction: $e_n$, a projector-like element connecting the last subsystem of the left side ($n$) with the first subsystem of the right side ($n+1$), bridging the wall.
    
    \item Right transpositions: $s_i$, with 
    $i=n+1,\ldots,n+m-1$, which correspond to swapping the $i$-th subsystem with the $(i+1)$-th subsystem within the last $m$ subsystems, i.e., right to the wall.
\end{itemize}

These generators satisfy the standard symmetric group relations within the left and right sectors, respectively, together with another relation between the two sectors as follows
\begin{subequations}
\begin{align}
    &s_i^2=1\,,\\
    &s_is_j=s_js_i\quad (|i-j|>1)\,,\\
    &s_is_{i+1}s_i=s_{i+1}s_is_{i+1}\,,\\
    &e_n^2=de_n\,,\label{subeq:pseudo-idempotent}\\
    &s_ie_n=e_ns_i\quad (i\neq n\pm1)\,,\\
    &e_ns_ie_n=e_n\quad (i=n\pm1)\,,\\
    &e_ns_{n+1}s_{n-1}e_ns_{n-1}=e_ns_{n+1}s_{n-1}e_ns_{n+1}\,,\\
    &s_{n-1}e_ns_{n+1}s_{n-1}e_n=s_{n+1}e_ns_{n+1}s_{n-1}e_n\,.
\end{align}    
\end{subequations}

In the context of quantum information, we have a clear connection with well-known operators. In the case of a simple transposition, $s_i$, it corresponds to the SWAP operator between the $i$-th and $(i+1)$-th systems
\begin{equation}
    s_i=\sum_{j,k=0}^{d-1}\ket{j,k}_{i,i+1}\bra{k,j}_{i,i+1}\,.
\end{equation}

Whereas a contraction element corresponds to the identity vectorized, i.e., the Choi matrix of the identity channel

\begin{equation}
    e_n=\iddk_{n,n+1}=\sum_{i,j=0}^{d-1}\ket{ii}_{n,n+1}\bra{jj}_{n,n+1}\,.
\end{equation}

The composition of two elements in the WBA can be diagrammatically represented through vertical concatenation. To compute the product of two elements $A_1A_2$, we place $A_1$ directly on top of $A_2$, identifying the bottom row of dots of the first element with the top row of dots of the second element.

Through this concatenation, we obtain a continuous set of connections between the top row of $A_1$ and the bottom row of $A_2$. This resulting configuration defines the new diagram, $A_3$.

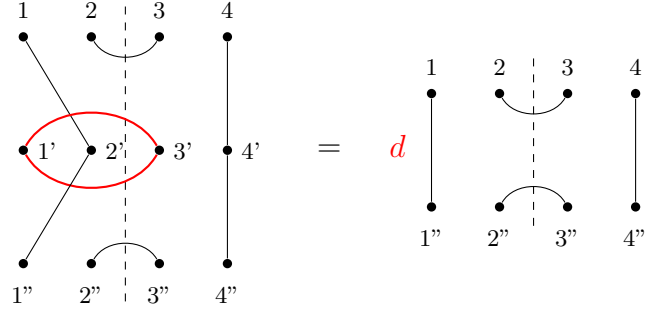
\begin{figure}[ht!]
    \centering
\begin{tikzpicture}[scale=1,
  every node/.style={circle, fill=black, minimum size=3.5pt, inner sep=0pt}]
  \def\sep{1.2*0.75}  
  \def\vertsep{2*0.75} 

  \foreach \i in {1,...,4} {
    \node[label={[label distance=4pt]above:{\i}}] (t\i) at (\i*\sep, 2*\vertsep) {};
  }
  \foreach \i in {1,...,4} {
    \node[label={[label distance=2pt]right:{\i'}}] (b\i) at (\i*\sep, \vertsep) {};
  }
  \foreach \i in {1,...,4} {
    \node[label={[label distance=4pt]below:{\i''}}] (bb\i) at (\i*\sep, 0) {};
  }
  
  \draw[dashed] (2.5*\sep, -0.5) -- (2.5*\sep, 2*\vertsep+0.5);

  \draw (t1) -- (b2);
  \draw (t2) to[out=-60, in=-120] (t3);
  \draw[red,thick] (b1) to[out=-300, in=-240] (b3);
  \draw (t4) -- (b4);

  \draw[red,thick] (b1) to[out=300, in=240] (b3);
  \draw (b2) -- (bb1);
  \draw (bb2) to[out=60, in=120] (bb3);
  \draw (b4) -- (bb4);

    \foreach \i in {1,...,4} {
    \node[label={[label distance=4pt]above:{\i}}] (tr\i) at (\i*\sep+6*\sep, 1.5*\vertsep) {};
  }
  \foreach \i in {1,...,4} {
    \node[label={[label distance=4pt]below:{\i''}}] (br\i) at (\i*\sep+6*\sep, 0.5*\vertsep) {};
  }
  \node[draw=none, fill=none] (const) at (5.5*\sep, \vertsep) {{\large \textcolor{black}{$=$}}};
    \node[draw=none, fill=none] (const) at (6.5*\sep, 1.04*\vertsep) {{\large \textcolor{red}{$d$}}};
    \draw (tr1) -- (br1);
  \draw (tr2) to[out=-60, in=-120] (tr3);
  \draw (br2) to[out=60, in=120] (br3);
  \draw (tr4) -- (br4);
\draw[dashed] (8.5*\sep, 0.5) -- (8.5*\sep, 1.5*\vertsep+0.5);
\end{tikzpicture}
    \caption[Composition of two elements of the walled Brauer algebra.]{An example of an element of the walled Brauer algebra, acting on $V^{\otimes 2}\otimes (V^*)^{\otimes 2}$. The dashed line separates the usual vector spaces from dual vector spaces.}
    \label{fig:wba-element-composition}
\end{figure}

In the resulting concatenation, internal closed loops may form exclusively within the middle section. These loops do not connect to the top or bottom dot rows of the composite diagram. For every such closed loop generated, the resulting diagram is multiplied by the parameter of the algebra, $d$ (in our case, the local dimension of the quantum systems). This relation arises from the pseudo-idempotent relation in the generators' relations in Eq.~\eqref{subeq:pseudo-idempotent}. Therefore, the diagrammatic composition rule is given by 
\begin{equation}
    d^{\,l}A_3=A_1A_2
\end{equation}
where $l$ is the total number of internal loops formed. An example of such composition rule is shown in Fig.~\ref{fig:wba-element-composition} above.

\subsection*{Mixed Schur-Weyl duality}
The importance of the WBA in our work stems from its role in the mixed Schur-Weyl duality. Standard Schur-Weyl duality establishes a connection between the actions of both the unitary, $\mathrm{SU}(d)$, and the symmetric group, $S_n$, on a Hilbert space of $n$ parties of local dimension $d$. The actions of both groups are simply to apply the same unitary transformation to each of the parties for the case of the unitary group and permuting the parties in the case of the symmetric. 

In contrast, when working with mixed Schur-Weyl duality, the collective action of the unitary group on the Hilbert space, $(\mathbb{C}^d)^{\otimes n}\otimes(\mathbb{C}^d)^{\otimes m}$, is given by the tensor product
\begin{equation}
    U^{\otimes n}\otimes {U^*}^{\otimes m}\,,
\end{equation}
where as in the main text $U^*$ is the complex conjugate of $U$.

This action commutes with the aforementioned action of the algebra of partially transposed permutations $\mathcal{A}_{n,m}^d$ on the Hilbert space. In other words, any operator, $A$, belonging to this algebra commutes with the mixed unitary action of $\mathrm{SU}(d)$
\begin{equation}
    [A,U^{\otimes n}\otimes {U^*}^{\otimes m}]\,,
\end{equation}
for all $U\in \mathrm{SU}(d)$.

Furthermore, as a consequence of this commutant relation, the Hilbert space can be decomposed into a direct sum of irreducible subspaces that are invariant under the action of $\mathrm{SU}(d)$ and $\mathcal{A}_{n,m}^d$, as 
\begin{equation}
    \mathcal{H}^{\otimes (n+m)}=\bigoplus_{\hat{\lambda}}\mathcal{H}_{\hat{\lambda}}\otimes \mathcal{H}^{\hat{\lambda}}\,,
\end{equation}
where $\hat{\lambda}$ labels the irreducible representations (irreps) of both $\mathrm{SU}(d)$ and $\mathcal{A}_{n,m}^d$. The subscripts and superscripts refer to the unitary group and algebra of partially transposed permutations, respectively, to resolve any ambiguity that may arise.

The basis in which the Hilbert space decomposes in the block-diagonal structure is called the \textit{Schur basis} and the basis change between the computational basis and the new one is called \textit{mixed Schur transform}, which has been recently studied in \cite{grinko_gelfand-tsetlin_2023}. In the Schur basis, any unitary transformation admits a representation in the irreducible subspaces as
\begin{equation}
    U^{\otimes n}\otimes (U^*)^{\otimes m}=\bigoplus_{\hat{\lambda}}U_{\hat{\lambda}}\otimes \mathbbm{1}^{\hat{\lambda}}\,,
\end{equation}
whereas any operator belonging to $\mathcal{A}_{n,m}^d$ can be expressed as
\begin{equation}
    A=\bigoplus_{\hat{\lambda}}\mathbbm{1}_{\hat{\lambda}}\otimes A^{\hat{\lambda}}\,.
\end{equation}

\subsection*{Irreducible representations and Young tableaux}

The irreps of both algebras are labeled by the same index, $\hat{\lambda}$. In the context of mixed Schur-Weyl duality, these labels correspond to pairs of partitions, $\hat{\lambda}=(\lambda_L,\lambda_R)$, which correspond to mixed Young diagrams. The left partition, $\lambda_{L}$
is associated with the $n$ subsystems to the left of the ``wall", while the right partition, $\lambda_{R}$, ¡
corresponds to the $m$ subsystems on the right.

\begin{figure}[ht!]
    \centering\vspace{0.5cm}
\begin{tikzpicture}[scale=0.6, every node/.style={font=\normalsize}]
  \node[draw=none, fill=none] (const) at (0,0) {\ydiagram{4,2}\quad \ \ydiagram{3,2,1}};
\end{tikzpicture}
    \caption{A mixed Young diagram, $\hat{\lambda}=(\lambda_L,\lambda_R)$ associated with an irrep of $\mathrm{SU}(d)$ and $\mathcal{A}_{6,6}^d$. In it, the total number of boxes are $\sum_{i}\lambda_{L,i}=\sum_{i}\lambda_{R,i}=6$. The number of rows are $\ell(\lambda_L)=2$, $\ell(\lambda_R)=3$. This mixed partition constitutes a valid irrep only for local dimensions $d\geq5$.}
    \label{fig:app_irrep_ex}
\end{figure}
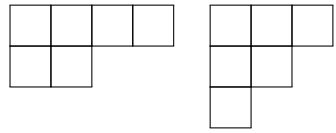

The total number of boxes in these partitions is given by a number, $q$, that satisfies $0\leq q \leq \min(n,m)$. Specifically, the sizes of the partitions are constrained to $\sum_{i}\lambda_{L,i}=n-q$ and $\sum_{i}\lambda_{R,i}=m-q$ for a given $q$. Furthermore, they are also restricted by the local dimension of the subsystems, $d$, which imposes a restriction on the total number of rows of both diagrams combined: $\ell({\lambda_L})+\ell(\lambda_R)\leq d$, where $\ell(\lambda)$ denotes the number of rows of a Young diagram. An example of a valid mixed Young diagram is shown in Fig.~\ref{fig:app_irrep_ex}. 

For a fixed local dimension $d$, these mixed Young diagrams can be mapped to standard Young diagrams. This mapping is achieved by embedding both left and right partitions into a single standard partition $\lambda$. We obtain the unified partition by concatenating the left partition with the complement of the right diagram with respect to its bounding rectangle:
\begin{equation}
    \lambda=(\lambda_{R,1}+\lambda_{L,1}-\lambda_{R,d}\,,\ \lambda_{R,1}+\lambda_{L,2}-\lambda_{R,d-1}\,,
    \ldots\,,\lambda_{R,1}+\lambda_{L,d-1}-\lambda_{R,2}\,,\lambda_{L,d} )\,.
\end{equation}

This approach to mixed Young diagrams allows us to use combinatorial formulae 
to compute the dimension of the irreducible subspace associated with $\hat{\lambda}$ corresponding to $\mathrm{SU}(d)$ via the Hook length formula or counting the number of semistandard Young diagrams~\cite{grinko_gelfand-tsetlin_2023}. 

\subsection*{Gelfand-Tsetlin basis}

To construct and work with operators within these irreducible subspaces, we require a basis in which the operators have the block diagonal form. To obtain such form, we use the Gelfand-Tsetlin basis, a canonical choice to construct the Schur basis systematically.

In the block-diagonal decomposition $\mathcal{H}^{\otimes(n+m)}=\bigoplus_{\hat{\lambda}}\mathcal{H}_{\hat{\lambda}}\otimes\mathcal{H}^{\hat{\lambda}}$, the Gelfand-Tsetlin basis provides an explicit set of orthonormal vectors that decomposes the global Hilbert spaces in this block-diagonal structure. Instead of working with operators $A\in \mathcal{A}_{n,m}^d$ represented on the global Hilbert space, we can work in the irreducible subspaces, $\hat{\lambda}$, of the algebra of partially transposed permutations by using this basis. This way, the dimensionality of the problem is drastically reduced: no degrees of freedom regarding the subspaces associated with the unitary group are taken into account. Any operator $A\in \mathcal{A}_{n,m}^d$ commutes with the mixed unitary action, Schur's Lemma dictates that it must act trivially on the unitary subsystem $\mathcal{H}_{\hat{\lambda}}$. Therefore, the operator is completely characterized by its matrix elements restricted to the subspaces associated with $\mathcal{A}_{n,m}^d$, that is, $\mathcal{H}^{\hat{\lambda}}$.

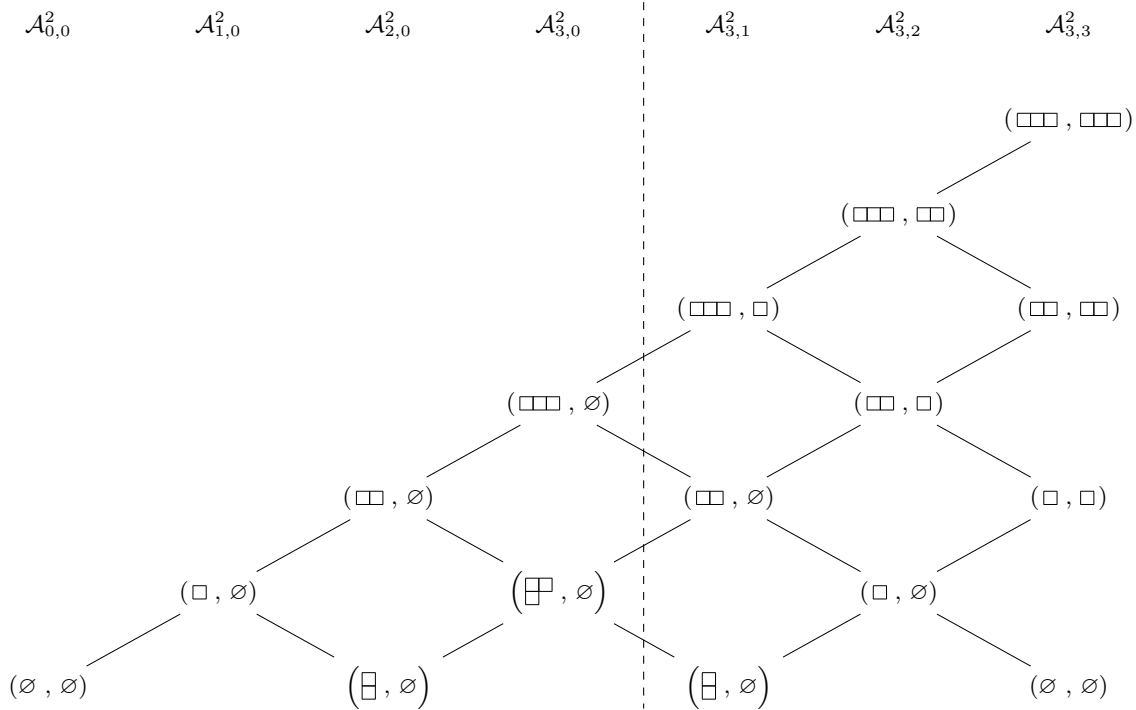
\begin{figure*}[!ht]
\centering
\begin{tikzcd}[every arrow/.append style={dash},column sep={2.25cm,between origins},row sep={1.25cm,between origins},
    execute at end picture={
        \draw[dashed] ([xshift=-5.625cm]\tikzcdmatrixname-1-7.north) -- ([xshift=-5.625cm]\tikzcdmatrixname-8-7.south);
    }]
\mathcal{A}_{0,0}^2&\mathcal{A}^2_{1,0}&\mathcal{A}^2_{2,0}&\mathcal{A}^2_{3,0}&\mathcal{A}^2_{3,1}&\mathcal{A}^2_{3,2}&\mathcal{A}^2_{3,3}\\
&&&&&&(\,\scalebox{0.35}{\ydiagram{3}}\ ,\,\scalebox{0.35}{\ydiagram{3}}\,)\ar[ld]\\
&&&&&(\,\scalebox{0.35}{\ydiagram{3}}\ ,\,\scalebox{0.35}{\ydiagram{2}} \,)\ar[ld]\ar[rd]& &\\
&&&&(\,\scalebox{0.35}{\ydiagram{3}}\ ,\,\scalebox{0.35}{\ydiagram{1}}\,) \ar[ld] \ar[rd]& &(\,\scalebox{0.35}{\ydiagram{2}}\ ,\,\scalebox{0.35}{\ydiagram{2}}\,)\ar[ld]\\
&&&(\,\scalebox{0.35}{\ydiagram{3}}\ ,\,\varnothing )\ar[ld] \ar[rd]&&(\,\scalebox{0.35}{\ydiagram{2}}\ ,\,\scalebox{0.35}{\ydiagram{1}}\,)\ar[ld]\ar[rd]&\\
&&(\,\scalebox{0.35}{\ydiagram{2}}\ ,\,\varnothing) \ar[ld] \ar[rd]&&(\,\scalebox{0.35}{\ydiagram{2}}\ ,\,\varnothing)\ar[rd] \ar[ld]&&(\,\scalebox{0.35}{\ydiagram{1}}\ ,\,\scalebox{0.35}{\ydiagram{1}}\,)\ar[ld]\\
&(\,\scalebox{0.35}{\ydiagram{1}}\ ,\,\varnothing )\ar[ld] \ar[rd]&&\Big(\hspace{0.03cm}\raisebox{0.1cm}{\scalebox{0.35}{\ydiagram{2,1}}}\ ,\,\varnothing\Big)\ar[ld]\ar[rd]&&(\,\scalebox{0.35}{\ydiagram{1}}\ ,\,\varnothing)\ar[ld]\ar[rd]&\\
(\varnothing\ ,\,\varnothing)&&\Big(\hspace{0.03cm}\raisebox{0.1cm}{\scalebox{0.35}{\ydiagram{1,1}}}\ ,\,\varnothing\Big)&&\Big(\hspace{0.03cm}\raisebox{0.1cm}{\scalebox{0.35}{\ydiagram{1,1}}}\ ,\,\varnothing\Big)&&(\varnothing\ ,\,\varnothing)
\end{tikzcd}
    \caption{Bratteli diagram for $n=m=3$ and $d=2$.}
    \label{fig:young-lattice}
\end{figure*}

Conceptually, the systematic construction of the Gelfand-Tsetlin basis relies on defining a chain of nested subalgebras and identifying the simultaneous eigenvectors of their invariant central elements through the paths of a Bratteli diagram, see Fig.~\ref{fig:young-lattice} below for an example of such diagrams in the context of partially transposed permutations. Instead of reproducing the extensive combinatorial and algebraic machinery required to build these states, we refer the interested reader to the comprehensive framework recently developed by Grinko \textit{et al.}~\cite{grinko_gelfand-tsetlin_2023} for the explicit formulation of the mixed Schur transform. From a practical perspective, to solve our specific optimization problem, it is sufficient to compute the matrix representations of the algebra generators ($s_i$ and $e_n$) exclusively within this basis. Because these generators span the entire algebra, any unitary covariant operator can be algebraically constructed directly from them. Consequently, the Choi matrices corresponding to our anomalous unitary hypotheses, as well as the optimal testers required to identify them, are constructed within the irreducible subspaces.

The complete algebraic procedure, including the explicit code used for the construction of the relevant operators and the proof for the two-anomaly case for qubits, is available in~\cite{llorens_Exact-identification-unknown-unitary} with some core functions adapted from~\cite{grinko_walledbrauer-gtbasis_2023}. 

\section{Local strategy in the general scenario}\label{app:local-strategy-general}

In this appendix, we detail the derivation of the success probability for the parallel local strategy in the general scenario involving $k\leq \floor{n/2}$ anomalies and arbitrary local dimension $d$. As introduced in the main text, due to the permutational symmetry of the hypotheses, the success probability is given by the expression 
\begin{equation}
    P_s(k;d) = \frac{1}{d^k}\mathrm{tr}\left(C^{(k)}\Pi_1^{\otimes k}\right)\,.
\end{equation}

As in the case of two anomalies, this expression can be reduced to an analytical expression. We achieve this by considering the two elements inside the trace. First, we expand the tensor product of projectors as an alternating sum 

\begin{equation}
    \Pi_1^{\otimes k} = \sum_{m=0}^k\frac{(-1)^m}{d^m}\Gamma_m\,,
\end{equation} 
where $\Gamma_m$ is the sum over all possible $m$-element subsets $I$, in which the projector $\Pi_0$ is acting  
\begin{equation}
    \Gamma_m = \sum_{\substack{I\subseteq\{1,2,\dots,k\} \\ |I|=m}}\bigotimes_{i\in I}\iddk_{i_{\mathrm{in}},i_{\mathrm{out}}}\,,
\end{equation}
where it is understood that the operator acts trivially on any subsystems outside $I$.

Second, we simplify the average Choi matrix $C^{(k)}$. Expressing this operator as matrices vectorized and the fact that the Haar measure is invariant under transposition, the Choi matrix from Eq.~\eqref{eq: General Choi} simplifies to a single Haar integral:
\begin{equation}
    C^{(k)} = \int|U\rangle\rangle\langle\langle U|^{\otimes k} dU\,.
\end{equation}

Substituting these two expressions into the expression for the success probability, we obtain
\begin{equation}
    P_s(k;d) = \frac{1}{d^k}\sum_{m=0}^k\frac{(-1)^m}{d^m}\int \mathrm{tr}\left(\Gamma_m|U\rangle\rangle\langle\langle U|^{\otimes k}\right) dU\,.
\end{equation}
Notice that the Choi matrix $C^{(k)}$ is invariant under any permutations of the $k$ bipartite systems. As a result, the trace of any individual term within $\Gamma_m$ depends solely on the number of non-trivial operators, $m$, and is entirely independent of the specific choice of the subset $I$. Summing over all such subsets yields a simple combinatorial coefficient $\binom{k}{m}$, leading to
\begin{equation}
\begin{aligned}
    \mathrm{tr}\left(\Gamma_m|U\rangle\rangle\langle\langle U|^{\otimes k}\right) = \binom{k}{m}\mathrm{tr}\left(\iddk^{\otimes m}|U\rangle\rangle\langle\langle U|^{\otimes k}\right) =\binom{k}{m}d^{k-m}\left|\langle\langle\mathbbm{1}|U\rangle\rangle\right|^{2m}\,.
\end{aligned}
\end{equation}

Substituting this result back into the sum, we obtain the final expression for the success probability:
\begin{equation}
P(k;d) = \frac{1}{d^{2k}}\sum_{m=0}^k(-1)^m\binom{k}{m}f_{m,d}\,d^{2(k-m)}\,,
\end{equation}
where the $f_{m,d}$ coefficients encode the group integral as
\begin{equation}
    f_{m,d}:= \int |\mathrm{tr}(U)|^{2m}dU\,.
\end{equation}

The coefficients $f_{m,d}$ generally do not have analytical expressions. However, in the cases where $d\geq m$, the coefficients count the number of permutations in $S_m$, and therefore $f_{m,d}=m!$. In general, $f_{m,d}$ counts the number of permutations of $S_m$ with longest increasing subsequence of length at most $d$, and can be computed from the dimensions $d_\lambda$ of the irreps of $S_m$~\cite{rains_increasing_1998}:
\begin{equation}
    f_{m,d} = \sum_{\substack{\lambda\vdash m \\ l(\lambda)\leq d}}d_\lambda^2\,.
\end{equation}

Finally, in the qubit scenario ($d=2$), these coefficients simplify to the Catalan numbers, $f_{m,d}=C_m$. This allows the expression for the success probability to be analytically evaluated, yielding the simple closed-form formula:
\begin{equation}
    P_s(k;d=2) = \frac{1}{2^{2k}}\binom{2k+1}{k+1}\,.
\end{equation}

\vfill\null

\end{document}